\documentclass[aps,prb,english,superscriptaddress,groupeaddress,twocolumn]{revtex4-1}
\usepackage[T1]{fontenc}
\usepackage[latin9]{inputenc}
\usepackage{color}
\usepackage{amsthm}
\usepackage{amsmath}
\usepackage{amssymb}
\usepackage{graphicx}

\makeatletter

\@ifundefined{textcolor}{}
{%
 \definecolor{BLACK}{gray}{0}
 \definecolor{WHITE}{gray}{1}
 \definecolor{RED}{rgb}{1,0,0}
 \definecolor{GREEN}{rgb}{0,1,0}
 \definecolor{BLUE}{rgb}{0,0,1}
 \definecolor{CYAN}{cmyk}{1,0,0,0}
 \definecolor{MAGENTA}{cmyk}{0,1,0,0}
 \definecolor{YELLOW}{cmyk}{0,0,1,0}
}

\makeatother

\usepackage{babel}
\begin{document}

\title{Majorana bound states and subgap states in three-terminal topological superconducting nanowire-quantum dot hybrid devices}

\author{Guang-Yao~Huang}
\affiliation{Beijing Key Laboratory of Quantum Devices, Key Laboratory for the
Physics and Chemistry of Nanodevices, and Department of Electronics,
Peking University, Beijing 100871, China}

\author{Xin Liu}
\affiliation{School of Physics and Wuhan National High Magnetic Field Center,
Huazhong University of Science and Technology, Wuhan, Hubei 430074,
China}

\author{H.~Q.~Xu}
\email[Corresponding author: ]{hqxu@pku.edu.cn; hongqi.xu@ftf.lth.se}
\affiliation{Beijing Key Laboratory of Quantum Devices, Key Laboratory for the
Physics and Chemistry of Nanodevices, and Department of Electronics,
Peking University, Beijing 100871, China}
\affiliation{Division of Solid State Physics, Lund University, Box 118, S-221
00 Lund, Sweden}

\date{\today}

\begin{abstract}
Three-terminal topological superconducting nanowire (TSNW)-quantum dot (QD) hybrid junction devices are studied. The energy spectra and the wave functions  of the subgap states
are calculated as a function of the superconducting phase
differences between TSNWs and as a function of the QD level energy based on the Bogoliubov-de Gennes tight-binding Hamiltonians. It is shown that when the QD level is located near or inside the superconducting gap, there can exist eight subgap states. Among them, four low energy (two positive and two negative) subgap states are essentially formed by linear combinations of the six Majorana bound states (MBSs) located at the ends of the three TSNWs. The remaining four high energy subgap states are mainly built from linear combinations of the QD state and the three MBSs of the TSNWs adjacent to the QD. When there is no QD level near or inside the superconducting gap, only six subgap states built from linear combinations of the six MBSs of the three TSNWs can be present in the system. It is also shown that there exists a unique point in the parameter space of the superconducting phase differences between TSNWs, at which the energies of the six low energy subgap states move close to each other towards nearly zero energies. Simple but general effective model Hamiltonians for the three-terminal TSNW-QD hybrid devices have also been developed. Based on the effective model Hamiltonians, the subgap states of the three-terminal TSNW-QD hybrid devices in the limit of the three infinitely long TSNWs are studied.  The results of the calculations and the effective model Hamiltonians could be used as a starting point to construct and  investigate the braiding schemes of MBSs in TSNW junction devices.

\end{abstract}
\maketitle

\section{Introduction}

Currently, Majorana bound states (MBSs) in condensed matter physics \cite{pu44.131,np5.614,rmp87.137,sst27.124003,nnano8.149,arcmp4.113,rpp75.076501}
attract great attention because of its non-Abelian statistics and potential applications in topological quantum computation
(TQC) \cite{rmp80.1083,ap303.2}. A series
of experiments towards detecting signatures of the MBSs have been performed on nanowire devices \cite{science336.1003,nl12.6414,np8.795,prb87.241401,sr4.7261,arxiv1603.04069,np8.887,prl110.126406,nature531.206,science354.1557,science346.602,npjqi2.16035,np13.286} made from semiconductors InSb and InAs nanowires, and ferromagnetic atomic chains. These nanowire systems are believed to be the most promising systems for generating MBSs. To realize MBSs with a semiconductor nanowire, it is required
that the nanowire should possess a large $g$-factor and strong spin-orbit
interaction (SOI). A topological superconducting nanowire with MBSs present at its two ends can be created by placing such a semiconductor nanowire in the proximity of an
$s$-wave superconductor and under application of a moderate magnetic
field \cite{prl105.077001,prl105.177002,prl106.127001,prb84.144522}.

Stimulated by the experimental progress \cite{science336.1003,nl12.6414,np8.795,prb87.241401,sr4.7261,arxiv1603.04069,np8.887,prl110.126406,nature531.206,science354.1557,science346.602,npjqi2.16035,np13.286}, many theoretical proposals
\cite{pra82.052322,prb84.035120,prb85.144501,np7.412,prb84.094505,prb91.174305,pra87.022343,njp14.035019,prb88.035121,prb91.201102,prb94.014511,prb94.035424,prx6.031016}
to implement TQC with MBSs in nanowire devices are reported. Because
braiding two MBSs, the essential operation to realize the TQC, is
not straightforward in one dimension, a solution to walk around
this difficulty is to build a nanowire network for MBS braiding or exchanging.
Before studying the actual physical process of MBS braiding in a nanowire network, the first natural step is to study the energy spectra of MBSs and subgap states in a minimal structure, a
three-terminal hybrid nanowire junction with the nanowires in the topological superconducting
phase. Various nanowire junctions \cite{nl8.1100,nnano8.859,nl13.5190,am26.4875,aem2.1500460,nl16.1933,nl17.531,nl17.2596,arXiv1703.05195,arXiv1705.01480}
have been realized and characterized by transport measurements. Several theoretical studies on multiterminal superconductor junctions have also been reported \cite{prb92.155437,prb92.205409,ncomms7.11167}.

In this paper, we report on a theoretical study of three-terminal topological superconducting nanowire (TSNW) junction devices with the nanowires connected to each other via a quantum dot (QD) in the junction region in each device. We explore the effects of the couplings among MBSs presented in the junction devices. Since, as shown in a recent work,\cite{prb95.235305} a QD can be an important constitution to realize
MBS braiding, here we include a QD in the devices. The QD is located at the connecting junction of the three nanowires and has an energy state which is assumed to be tunable using, e.g., a gate. Thus, the QD can play a role of high barrier (without
a QD level at the Fermi energy) or a conduction channel (with a QD
level at the Fermi energy) and can therefore be used to control the couplings among MBSs in the TSNWs. Our three-terminal TSNW junction devices are described by standard Bogoliubov-de Gennes (BdG) tight-binding Hamiltonians. The energy spectra and the probability distributions of the wave functions
of subgap states in the devices are calculated and analyzed. We have also extracted low-energy effective models that describe the results of the numerical calculations for the subgap states of the three-terminal TSNW-QD hybrid devices based on the BdG tight-binding Hamiltonians. We demonstrate that the coupling strengths between the MBSs adjacent to the QD in the three-terminal TSNW-QD hybrid devices are functions of the
azimuths of the TSNWs, the QD  level energy, and the superconducting phase differences between the TSNWs. Tuning superconducting phase differences could be achieved by employing local magnetic fields to SQUID setups connecting the TSNWs.  A similar but different all-magnetic-flux-controlled prototype device for MBS braiding is proposed in Refs.~\onlinecite{njp14.035019,prb88.035121},
in which the magnetic flux is used to control the coupling of MBSs in a single
nanowire. Here, the magnetic flux is used to tune the couplings between MBSs in
different nanowires.

The paper is organized as follows. In Sec. II, we present the BdG tight-binding model that is used to describe our three-terminal TSNW-QD hybrid junction devices.
In Sec. III, we present the results of the calculations for the three-terminal TSNW-QD hybrid junction devices based on the BdG tight-binding model. The energy spectra and wave functions of the subgap states are calculated in the parameter space of the phase differences between TSNWs for two cases, i.e., the case when there is no QD level located near or inside the superconducting gap and the case when the QD level is located near or inside the superconducting gap. Simple but general effective models extracted for describing the subgap states of the three-terminal TSNW-QD hybrid devices in the two cases will also be presented and discussed. Finally, in Sec. IV, we summarize and conclude the paper.

\section{Model and formalism}

\begin{figure}
\centering\includegraphics[scale=0.5]{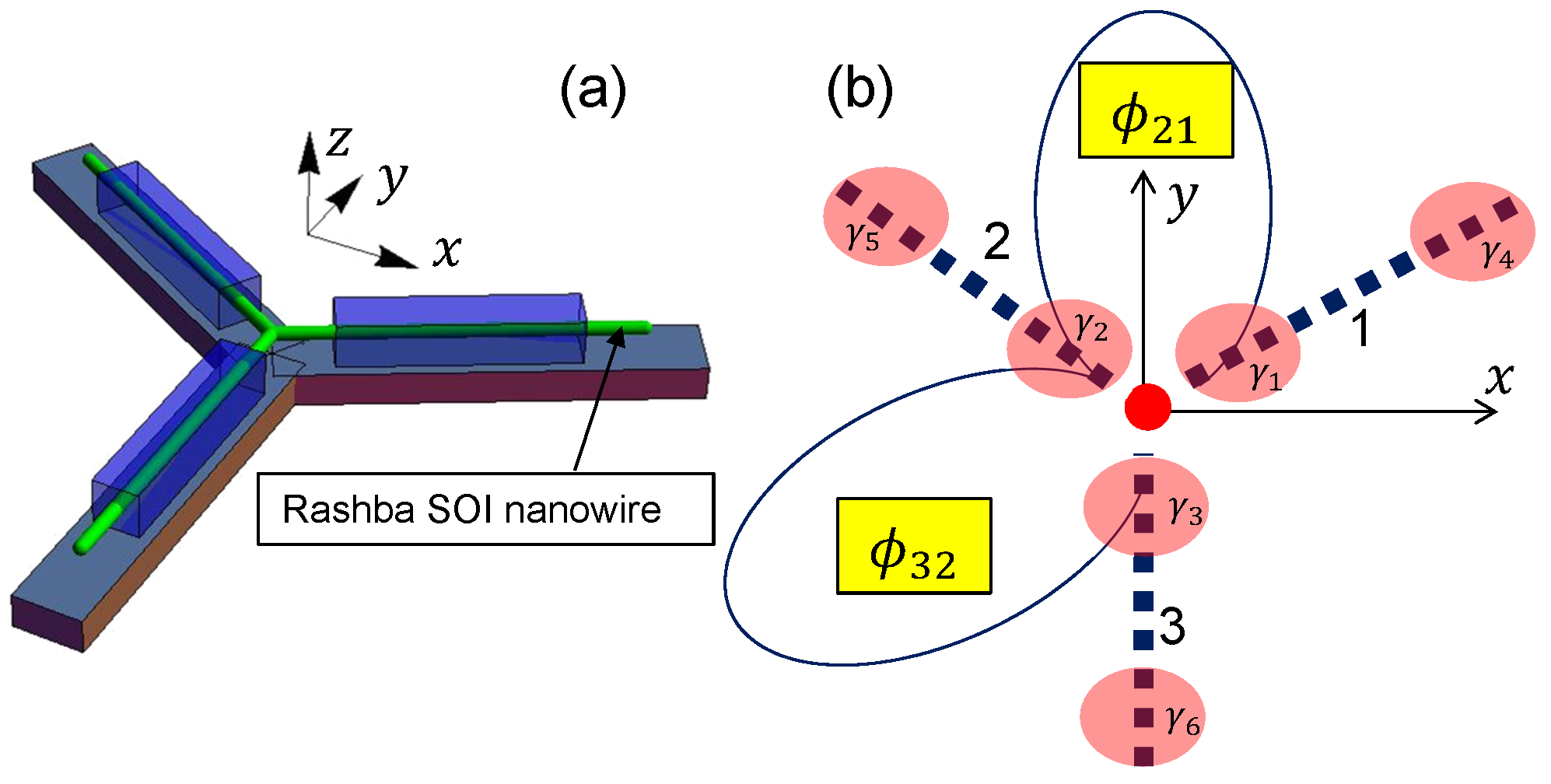}\protect\caption{\label{setup}(a) Sketch of a three-terminal TSNW-QD hybrid device.
Three semiconductor nanowires, coupled by a QD in the center, are placed on the $x$-$y$ plane and covered by $s$-wave
superconductors. The nanowires are assumed to possess strong Rashba spin-orbit interaction and a magnetic field is assumed to apply to the device along the $z$ direction. (b) Simplified view of the device. The QD is
represented by a red dot in the origin of the coordinate system,  $\gamma_{i}$ ($i=1,\dots,6$)
label the MBSs (marked by pink color shaded circles) at the ends of the TSNWs,  $\phi_{21}$ and
$\phi_{32}$ are the superconductor phase differences between TSNWs 1 and 2 and between TSNWs  2 and 3. Note that $\phi_{21}$ and
$\phi_{32}$ can be tuned with magnetic fluxes through properly designed SQUID structures as indicated in the figure. The orientations of the three TSNWs are defined by azimuths $q_1$, $q_2$, and $q_3$. In most results presented in this work, the systems with
$q_{1}=\frac{\pi}{6}$, $q_{2}=\frac{5\pi}{6}$, and $q_{3}=\frac{3\pi}{2}$ as drawn in the figure are considered. However, the results are easily generalized to a system with arbitrary values of $q_1$, $q_2$, and $q_3$.}
\end{figure}

As depicted in Fig.~\ref{setup}(a), we consider a three-terminal TSNW-QD hybrid junction device which is placed on the
$x$-$y$ plane with three TSNWs connected together via a
QD at the junction. An applied magnetic field along the $z$ direction is assumed
to drive the three semiconductor nanowires with strong SOI and a sizable $g$ factor into the topological superconducting phase. Figure \ref{setup}(b) displays a lattice model of the system with the TSNWs
represented by discrete sites and the QD by one site at the center of the junction. The two of the TSNWs are assumed to be connected via superconductor leads and
$\phi_{ij}$ is the phase difference between TSNWs $i$ and $j$, which
can be tuned by applying a local magnetic field. The coordinate system is established by taking
the QD site as the origin.

The total Hamiltonian includes three parts and can be written as
\begin{equation}
H_{total}=\sum_{i=1}^{3}H_{i}+H_{QD}+H_{c},
\end{equation}
where $H_{i}$ is the Hamiltonian of the $i$-th TSNW, $H_{QD}$ is the Hamiltonian of the QD in normal phase (i.e., without including
a superconducting pairing potential), and $H_{c}$ describes the couplings between
the TSNWs and the QD. In order to take into account the spatial directions of
the wires, the azimuth of the $i$-th TSNW $q_{i}$\textcolor{blue}{{}
}enters $H_{i}$ following the formula developed in Ref.~\onlinecite{prb85.144501}. In the tight-binding approach, the Hamiltonian $H_{i}$ takes the form of
\begin{equation}
H_{i}=\sum_{n,m}\frac{1}{2}\Psi_{n}^{\dagger}(\mathcal{H}_{BdG}^{i})_{n,m}\Psi_{m},
\end{equation}
where $\mathcal{H}_{BdG}^{i}$ is the BdG Hamiltonian
for the $i$-th TSNW in the Nambu spinor basis of $\Psi_{n}=[\psi_{n\uparrow},\psi_{n\downarrow},\psi_{n\downarrow}^{\dagger},-\psi_{n\uparrow}^{\dagger}]^{T}$,
\begin{equation}
(\mathcal{H}_{BdG}^{i})_{n,m}=h_{0}^{i}\delta_{n,m}+h_{1}^{i}\delta_{n,m-1}+h_{-1}^{i}\delta_{n,m+1},
\end{equation}
with
\begin{eqnarray}
h_{0}^{i} &\! = \!& (2t-\mu)\tau_{z}+V_{z}\sigma_{z}+\Delta(\tau_{x}\cos\phi_{i}-\tau_{y}\sin\phi_{i}),\;\; \\
h_{1}^{i} &\!= \! & [-t+i\alpha_{0}(\sigma_{y}\cos q_{i}-\sigma_{x}\sin q_{i})]\tau_{z},\\
h_{-1}^{i} &\! =\!  & [-t-i\alpha_{0}(\sigma_{y}\cos q_{i}-\sigma_{x}\sin q_{i})]\tau_{z}.
\end{eqnarray}
Here, $m$ and $n$ are the site indices, $t=\frac{\hbar^{2}}{2m^{*}a^{2}}$
is the hopping parameter, which is related to the band width, with $\hbar$ standing for the reduced
Planck constant, $m^{*}$ the effective mass, and $a$ the lattice
spacing, $\sigma_{u}$ and $\tau_{u}$, where $u=x,y$ or $z$,  are the Pauli matrices acting, respectively,
on spin and particle-hole spaces,  $\alpha_{0}$ is the Rashba SOI strength,
$\mu$ is the chemical potential, $V_{z}$ is the Zeeman energy, $\Delta$ and $\phi_{i}$
are the magnitude and phase of the superconducting pairing potential, and $q_{i}$ is the
azimuth of the $i$-th TSNW.

The topological phase transition point for each nanowire is at $V_{z}=\sqrt{\Delta^{2}+\mu^{2}}$,
with $V_{z}<\sqrt{\Delta^{2}+\mu^{2}}$ for the nanowire in trivial phase and $V_{z}>\sqrt{\Delta^{2}+\mu^{2}}$
for the nanowire in topological superconducting phase \cite{prl105.077001,prl105.177002}.
For simplicity, in the calculations and discussions presented below, the chemical
potential $\mu$ is set to 0 and $V_{z}=2\Delta$, such that all the nanowires
are in topological superconducting phase.

The Hamiltonian of the QD in the basis of $\Psi_{QD}=[\psi_{QD\uparrow},\psi_{QD\downarrow},\psi_{QD\downarrow}^{\dagger},-\psi_{QD\uparrow}^{\dagger}]^{T}$
is
\begin{equation}
H_{QD}=\frac{1}{2}\Psi_{QD}^{\dagger}\mathcal{H}_{QD}\Psi_{QD},
\end{equation}
with
\begin{equation}
\mathcal{H}_{QD}=-eV_{g}\tau_{z}+V_{z}\sigma_{z},
\end{equation}
where $-e$ is the single electron charge and $V_{g}$ is the tunable
gate voltage.

The coupling Hamiltonian is
\begin{equation}
H_{c}=\sum_{i=1}^{3}\frac{1}{2}\Psi_{i}^{\dagger}T_{i}\Psi_{QD}+\mathrm{H.c.},
\end{equation}
where $\Psi_{QD}$ and $\Psi_{i}$ are the annihilation operators
at the QD site and its adjacent site in the $i$-th TSNW, and $T_{i}=-t_{ci}\tau_{z}$
is the spin independent coupling matrix.

\section{Subgap states and effective Hamiltonians}

As can be seen in the above formulation, the energy spectra and the coupling between MBSs adjacent to the QD are functions of the phase differences, $\phi_{21}=\phi_2-\phi_1$ and $\phi_{32}=\phi_3-\phi_2$ between TSNWs, and the energy position of the QD level. In the following, we first in Sec. III A present the results of the calculations for the case when there is no QD level in or near the superconducting gap. Here, we will show that the MBSs adjacent to the QD and the couplings between them can be tuned through tuning phase differences $\phi_{21}$ and $\phi_{32}$. In Sec. III B, we will present the results of calculations for the case when a QD level is present in or near the superconducting gap and show that the tunings of the MBSs and the couplings between them can be also achieved through controlling of the energy level in the QD. Simple effective model Hamiltonians for describing the results of the numerical calculations will also be extracted for the two cases. A brief discussion about the generality of the effective models will be given in Sec. III C.

\subsection{System with no QD level in and near the superconducting gap}

In a three terminal TSNW-QD hybrid system with all QD levels being located far from the superconducting gap and, thus, far from the Fermi
level, the QD just acts as a potential barrier at the connection point of
the three TSNWs. Below, we will present the results of calculations for the energy spectra and the wave functions of the subgap states of the system by fixing $\phi_{21}$ while varying $\phi_{32}$. If the calculations are performed by fixing $\phi_{32}$ and varying $\phi_{21}$, the same results will be obtained.

\begin{figure*}
\centering\includegraphics[scale=0.8]{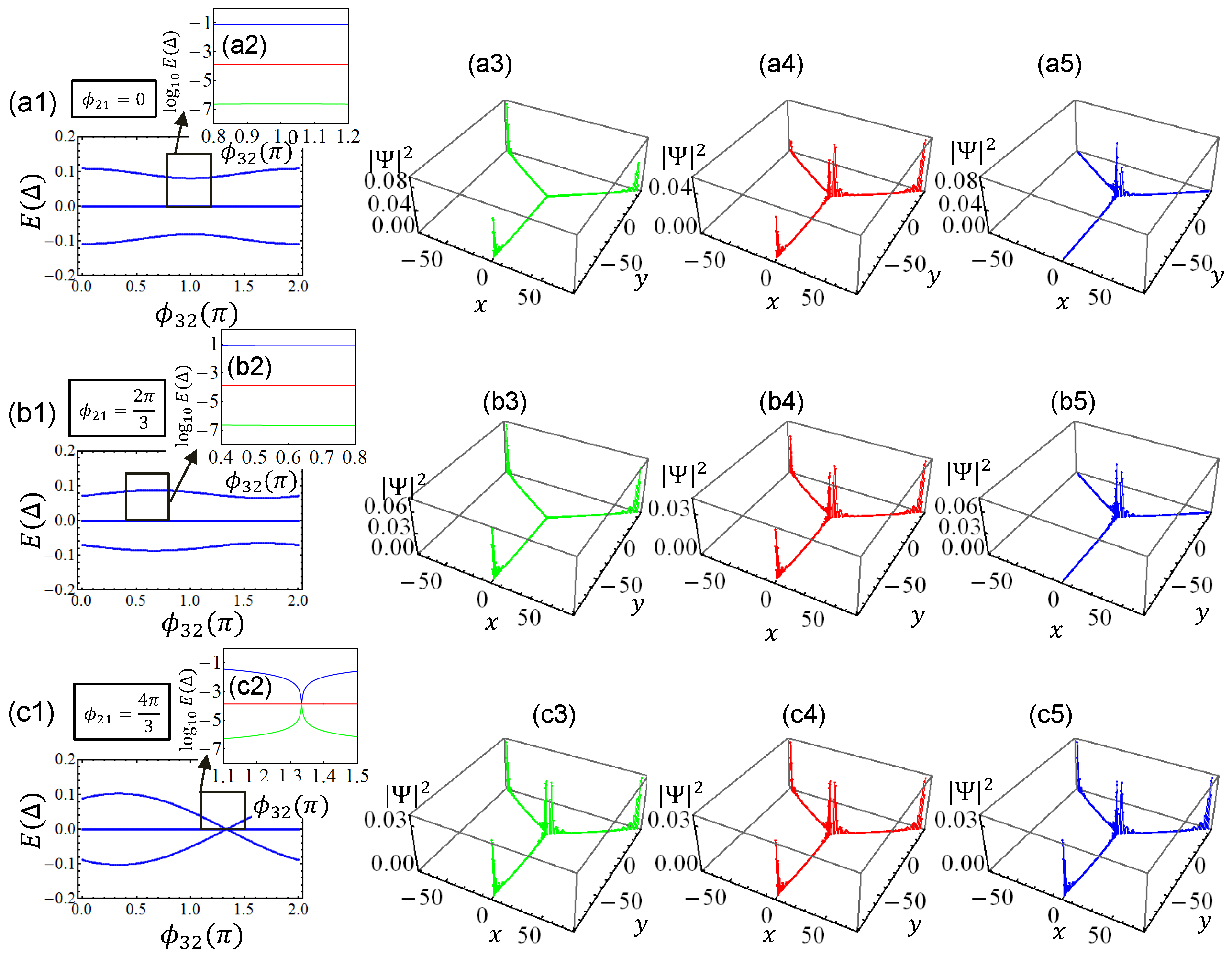}\protect\caption{\label{TBEnAndWaveFuncWithoutResonantLevel}
Subgap state energy spectra and wave functions of a three-terminal TSNW-QD hybrid system with no QD level near or inside the superconducting gap calculated based on the full BdG tight-binding Hamiltonian. (a1) Subgap state energy spectra calculated for the system with phase difference $\phi_{21}=0$ against phase difference $\phi_{32}$. (a2) Zoom-in view of the energy spectra of the three positive energy subgap states of the system in a logarithmic scale in the region marked by a rectangle in (a1). (a3)-(a5) Probability distributions $|\Psi_{n}|^{2}$ of the three positive energy subgap states calculated for the system with $\phi_{21}=0$ and $\phi_{32}=\pi$. (b1)-(b2) The same as (a1)-(a2) but for $\phi_{21}=\frac{2\pi}{3}$. (b3)-(b5) The same as (a3)-(a5) but for $\phi_{21}=\frac{2\pi}{3}$ and $\phi_{32}=\frac{3\pi}{5}$.
(c1)-(c2) The same as (a1)-(a2) but for $\phi_{21}=\frac{4\pi}{3}$. (c3)-(c5) The same as (a3)-(a5) but for $\phi_{21}=\phi_{32}=\frac{4\pi}{3}$. Other
parameters employed in the calculations are $V_{z}=2\Delta$, $\mu=0$, $t=10\Delta$, $\alpha_{0}=2\Delta$, and $t_{ci}=3\Delta$ (with $i=1,2,\mbox{and}\; 3$). Each nanowire is modeled with 90 tight-binding sites.}
\end{figure*}

Figure \ref{TBEnAndWaveFuncWithoutResonantLevel} shows the calculated energy spectra of subgap state of the system at different values of $\phi_{21}$ and $\phi_{32}$. In Figs.~\ref{TBEnAndWaveFuncWithoutResonantLevel} (a1) and \ref{TBEnAndWaveFuncWithoutResonantLevel}(a2), the subgap energy spectra of the system with $V_z=2\Delta$ at a fixed value of $\phi_{21}=0$ and different values of $\phi_{32}$ are plotted in a linear and a logarithmic scale, respectively. There are six subgap states
inside the superconducting gap. These states appear in pairs at energies of $\sim\pm10^{-7}\Delta$, $\sim\pm10^{-4}\Delta$ and $\sim\pm10^{-1}\Delta$, see Fig.~\ref{TBEnAndWaveFuncWithoutResonantLevel}(a2) for the three positive energy subgap states. Note that at $V_{z}=2\Delta$, the superconducting gap parameter $E_g$ shrinks to $E_{g}\approx0.7\Delta$. The wave functions of the three positive energy subgap states calculated at $\phi_{21}=0$  and $\phi_{32}=\pi$  are shown in Figs.~\ref{TBEnAndWaveFuncWithoutResonantLevel}(a3) to \ref{TBEnAndWaveFuncWithoutResonantLevel}(a5). The probability distribution of the lowest positive energy subgap state at the energy of $\sim 10^{-7}\Delta$ is
shown in Fig.~\ref{TBEnAndWaveFuncWithoutResonantLevel}(a3). It is seen that this state
is formed from the three MBSs located at the outer
ends of the three TSNWs. The probability distribution of the next lowest positive energy subgap state at the energy of $\sim 10^{-3}\Delta$ is shown in Fig.~\ref{TBEnAndWaveFuncWithoutResonantLevel}(a4). This state is formed from a superposition of all the six MBSs. Note that the energies of these two states are very close to zero, but are finite due to the fact that the TSNWs are all finite in length. The probability distribution of the remaining positive energy subgap state at the energy of $\sim 10^{-1}\Delta$ is shown in Fig.~\ref{TBEnAndWaveFuncWithoutResonantLevel}(a5). This state is formed from the hybridization of the MBSs at the inner ends of the three TSNWs. Thus, the energy of this state would not change with increasing TSNW lengths. However, the state can be effectively tuned by
the couplings $t_{ci}$ between the TSNWs and the QD, and by the phase differences between the TSNWs, as we will show below. The three corresponding negative energy subgap states show the same probability distributions as their positive energy counterparts.

Figures \ref{TBEnAndWaveFuncWithoutResonantLevel}(b1) and \ref{TBEnAndWaveFuncWithoutResonantLevel}(b2) show the subgap energy spectra of the system at a fixed value of $\phi_{21}=\frac{2\pi}{3}$ and different values of $\phi_{32}$.
Figures \ref{TBEnAndWaveFuncWithoutResonantLevel}(b3) to \ref{TBEnAndWaveFuncWithoutResonantLevel}(b5) show the wave functions of the subgap states of the system at $\phi_{21}=\frac{2\pi}{3}$ and $\phi_{32}=\frac{3\pi}{5}$. Clearly, the wave functions of the six subgap states show the similar
probability distributions as their corresponding subgap states shown in Figs. \ref{TBEnAndWaveFuncWithoutResonantLevel}(a3) to \ref{TBEnAndWaveFuncWithoutResonantLevel}(a5). In fact, similar characteristics in the probability distributions of the six subgap states are observed at other values of $\phi_{21}$ and $\phi_{32}$, except for the case of $\phi_{21}=\frac{4\pi}{3}$ and $\phi_{32}=\frac{4\pi}{3}$ as we will show below.

Figures \ref{TBEnAndWaveFuncWithoutResonantLevel}(c1) and \ref{TBEnAndWaveFuncWithoutResonantLevel}(c2) show the subgap energy spectra of the system at a fixed value of $\phi_{21}=\frac{4\pi}{3}$. Here, it is seen that the
energies of the subgap states are varied with varying $\phi_{32}$ and all the subgap states move together in energy when the phase difference $\phi_{32}$ moves to $\phi_{32}=\frac{4\pi}{3}$. We note that this particular case is found only when $\phi_{21}=\frac{4\pi}{3}$. We note also that at this particular case, all the subgap states are close to zero in energy as shown in Figs.~\ref{TBEnAndWaveFuncWithoutResonantLevel}(c1) and \ref{TBEnAndWaveFuncWithoutResonantLevel}(c2)
and have the same form of the probability distributions as seen in
Figs.~\ref{TBEnAndWaveFuncWithoutResonantLevel}(c3) to \ref{TBEnAndWaveFuncWithoutResonantLevel}(c5). It can also be observed that the subgap states within
each TSNW have the same form of probability distributions as that in an isolated single TSNW, which implies that
the couplings between the three TSNWs become vanishingly small at this specific case. These results indicate that the phase differences between the TSNWs can play a role in tuning of couplings between the three TSNWs and we
can control the couplings between the TSNWs via controlling phase differences between the three TSNWs.

\begin{figure*}
\centering\includegraphics[scale=0.8]{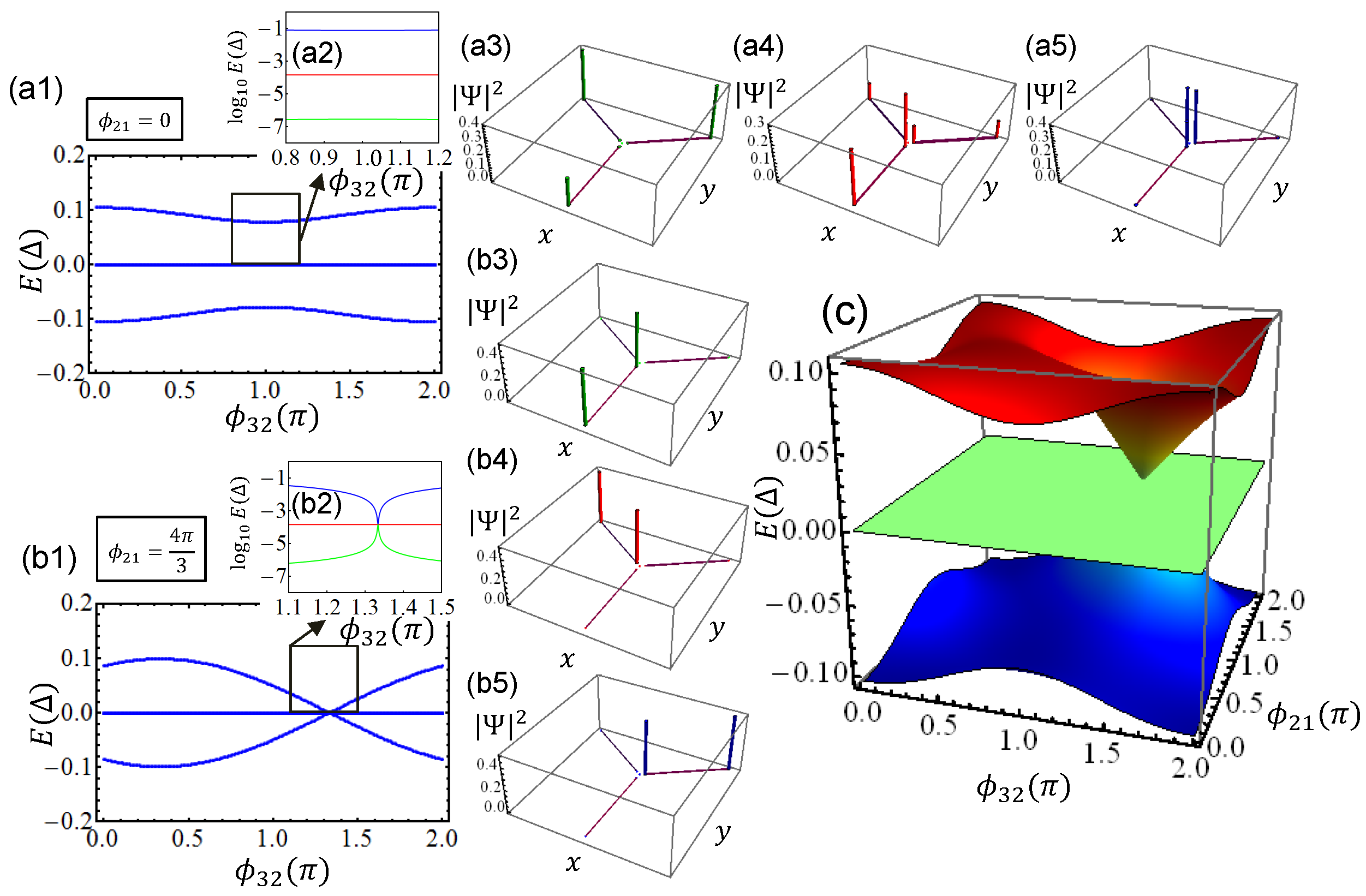}\protect\caption{\label{energyEffective}
Subgap state energy spectra and wave functions of a three-terminal TSNW-QD hybrid system with no QD level near or inside the superconducting gap calculated based on the effective model Hamiltonian of Eq.~(\ref{eqEffectiveWL}). (a1) Subgap state energy spectra calculated for the system with phase difference $\phi_{21}=0$ against phase difference $\phi_{32}$.  (a2) Zoom-in view of the energy spectra of the three positive energy subgap states of the system in a logarithmic scale in the region marked by a rectangle in (a1). (a3)-(a5) Probability distributions $|\Psi_{n}|^{2}$ of the three positive energy subgap states calculated for the system with $\phi_{21}=0$ and $\phi_{32}=\pi$. (b1)-(b2) The same as (a1)-(a2) but for $\phi_{21}=\frac{4\pi}{3}$. (b3)-(b5) The same as (a3)-(a5) but for $\phi_{21}=\phi_{32}=\frac{4\pi}{3}$. (c) Subgap state energy spectra of the system in the full parameter space of $\phi_{32}$ and $\phi_{21}$. The coupling parameters employed in the calculations are $t_{14}=t_{25}=t_{36}=0.0003\Delta$ and $t_{12}=t_{23}=t_{13}=0.14\Delta$.}
\end{figure*}

Based on the numerical results presented above, an effective model involving only the MBSs
in the three-terminal TSNW-QD hybrid structure can be constructed and can be used for simplification in considering employing the MBSs in the system. Similar to the procedure employed for two MBSs coupled
through a weak link in a one-dimensional topological superconducting system \cite{np7.412}, the effective Hamiltonian of the three-terminal TSNW-QD hybrid structure can be written as
\begin{equation}
H_{e\!f\!f}=H_{e\!f\!f}^{S}+H_{e\!f\!f}^{D},\label{eqEffectiveWL}
\end{equation}
with
\begin{eqnarray}
H_{e\!f\!f}^{S} & = & i(t_{14}\gamma_{1}\gamma_{4}+t_{25}\gamma_{2}\gamma_{5}+t_{36}\gamma_{3}\gamma_{6}),\\
H_{e\!f\!f}^{D} & = & i(\gamma_{2}\gamma_{1}t_{21}\sin\frac{\beta_{21}}{2}+\gamma_{3}\gamma_{2}t_{32}\sin\frac{\beta_{32}}{2}\nonumber \\
 &  & +\gamma_{1}\gamma_{3}t_{13}\sin\frac{\beta_{13}}{2}),
\end{eqnarray}
where $t_{ij}\in\mathbb{R}$ denotes the coupling between MBSs $\gamma_{i}$
and $\gamma_{j}$ [see the schematic in Fig. \ref{setup}(b) for the locations of the MBSs] and satisfies $t_{ij}=-t_{ji}$, and phase difference $\beta_{ji}$ is defined as $\beta_{ji}\equiv\beta_{j}-\beta_{i}=q_{j}-q_{i}+\phi_{j}-\phi_{i}$ and can be written explicitly as
\begin{eqnarray}
\beta_{21} & = & \phi_{21}+q_{2}-q_{1},\nonumber \\
\beta_{32} & = & \phi_{32}+q_{3}-q_{2},\label{eqPhaseDifference}\\
\beta_{13} & = & -\beta_{21}-\beta_{32}.\nonumber
\end{eqnarray}
For an arbitrary junction, the azimuths $q_{1,2,3}$ are in the range $[0,2\pi]$. For the specific case drawn in Fig.~\ref{setup}(b), $q_1=\frac{\pi}{6}$, $q_2=\frac{5\pi}{6}$, and $q_3=\frac{3\pi}{2}$, therefore
\begin{eqnarray}
\beta_{21} & = & \phi_{21}+\frac{2\pi}{3},\nonumber \\
\beta_{32} & = & \phi_{32}+\frac{2\pi}{3},\label{eqPhaseDifference}\\
\beta_{13} & = & -\phi_{21}-\phi_{32}-\frac{4\pi}{3}.\nonumber
\end{eqnarray}

There are two types of couplings between MBSs. One contains the couplings between two MBSs in the same TSNW ($H_{e\!f\!f}^{S}$) and the other one contains the couplings between MBSs adjacent to the QD but belonging to different TSNWs ($H_{e\!f\!f}^{D}$). Comparing with
the couplings ($t_{ij}$) between MBSs in $H_{e\!f\!f}^{S}$, the couplings
($t_{ij}\sin\frac{\beta_{ji}}{2}$) in $H_{e\!f\!f}^{D}$ has two parts--a constant amplitude part $t_{ij}$ and an oscillatory part arising from
the phase difference between TSNWs. In contrast to a straight nanowire system with the two inner MBSs coupled with a weak link, where the phase difference
is solely determined by the magnetic flux through a circuit loop, the phase differences between different TSNWs
in the three-terminal TSNW-QD hybrid system could contain both terms induced by the magnetic fluxes and terms related to the azimuths of the TSNWs. The azimuths of the TSNWs enter the Hamiltonian due to the presence of SOI in the nanowires
\cite{prb85.144501}. It has been shown for a continuous system that
when a one-dimensional spinful Rashba $s$-wave superconducting nanowire model
is mapped to a spinless $p$-wave superconducting nanowire model, the effective phase of the system is given by the summation of the azimuth and the phase in the
$s$-wave superconducting pairing potential of the nanowire. Thus, the differences in these two parts constitute the total phase differences between the TSNWs.

Figure \ref{energyEffective} shows the results of calculations based on the effective Hamiltonian. It is seen that the results of the full numerical calculations based on the BdG tight-binding Hamiltonian are recovered by the calculations based on the effective Hamiltonian. In particular, the energy spectra and the probability distributions seen in Figs.~\ref{TBEnAndWaveFuncWithoutResonantLevel}(a1) to \ref{TBEnAndWaveFuncWithoutResonantLevel}(a5) are well reproduced by the calculations based on the effective model shown in Figs.~\ref{energyEffective}(a1) to \ref{energyEffective}(a5). Here, we should note that the wave function of each MBS obtained based on the effective Hamiltonian is a point-like function, while the wave function of a MBS calculated based on the BdG tight-binding model exhibits a spatial but strongly  localized distribution.
Figures \ref{energyEffective}(b1) to \ref{energyEffective}(b5) show the calculated
energy spectra for the case of $\phi_{21}=\frac{4\pi}{3}$ and the probability distributions for the case of $\phi_{21}=\phi_{32}=\frac{4\pi}{3}$ based on the effctive Hamiltonian, corresponding to the results calculated based on the BdG tight-binding Hamiltonian shown in Figs.~\ref{TBEnAndWaveFuncWithoutResonantLevel}(c1) to \ref{TBEnAndWaveFuncWithoutResonantLevel}(c5).
Here, it is seen that at $\phi_{21}=\phi_{32}=\frac{4\pi}{3}$ where the six MBSs become nearly degenerate, the three TSNWs are decoupled in the effective Hamiltonian and each MBS resembles exactly that in
an isolated TSNW. Note that these states are still
superpositions of MBSs within each TSNW, since the coupling of the MBSs
within each TSNW remains non-zero in order to simulate the finite length effect in
our effective model. Figure \ref{energyEffective}(c) shows a view of the energy spectra of the three-terminal TSNW-QD hybrid system in the whole
parameter space of $\phi_{21}$ and $\phi_{32}$ based on the effective Hamiltonian.

If the three TSNWs are semi-infinitely long, the overlap of two MBSs within each
TSNW can be neglected, the effective Hamiltonian can be reduced to
\begin{eqnarray}
H_{S\!I\!n\!f} & = & i(\gamma_{2}\gamma_{1}t_{21}\sin\frac{\beta_{21}}{2}+\gamma_{3}\gamma_{2}t_{32}\sin\frac{\beta_{32}}{2}\nonumber \\
 & + & \gamma_{1}\gamma_{3}t_{13}\sin\frac{\beta_{13}}{2}).
\label{HSInf}
\end{eqnarray}
In this limit, the three-terminal TSNW-QD hybrid structure consists of four zero-energy MBSs with three of them located at the outer ends of the three TSNWs and the remaining one in the vicinity of the QD.\cite{njp14.035019,prb85.144501} There are also two finite energy states in the system located in the vicinity of the QD. These two states originate from hybridization of the MBSs adjacent to the QD. Explicitly, based on the Hamiltonian $H_{S\!I\!n\!f}$ given in Eq.~(\ref{HSInf}), the eigenenergies of the three states localized in the vicinity of the QD can be readily obtained as
\begin{equation}
E_{0} = 0 \;\; \mbox{and}\;\; E_{\pm}  =  \pm\frac{\displaystyle s_{0}}{\displaystyle 2},
\end{equation}
where $s_{0}=\sqrt{s_{12}^{2}+s_{23}^{2}+s_{13}^{2}}$ with $s_{ji}=t_{ji}\sin\frac{\beta_{ji}}{2}$. The corresponding eigenvectors are
\begin{eqnarray}
|0\rangle &=& \frac{1}{s_{0}}(s_{32}\gamma_1+s_{13}\gamma_2+s_{21}\gamma_3),\\
|E_{\pm}\rangle  &=&  \frac{1}{s_{N}}[
(s_{13}s_{32}\pm is_{21}s_{0})\gamma_1
-(s_{21}^{2}+s_{32}^{2})\gamma_2 \nonumber
\\&+&(s_{21}s_{13}\mp is_{32}s_{0})\gamma_3],
\end{eqnarray}
where the normalized constant $s_{N}=s_{0}\sqrt{2(s_{21}^{2}+s_{32}^{2})}$.

\subsection{System with a QD level inside or near the superconducting gap}

\begin{figure}
\centering\includegraphics[scale=0.76]{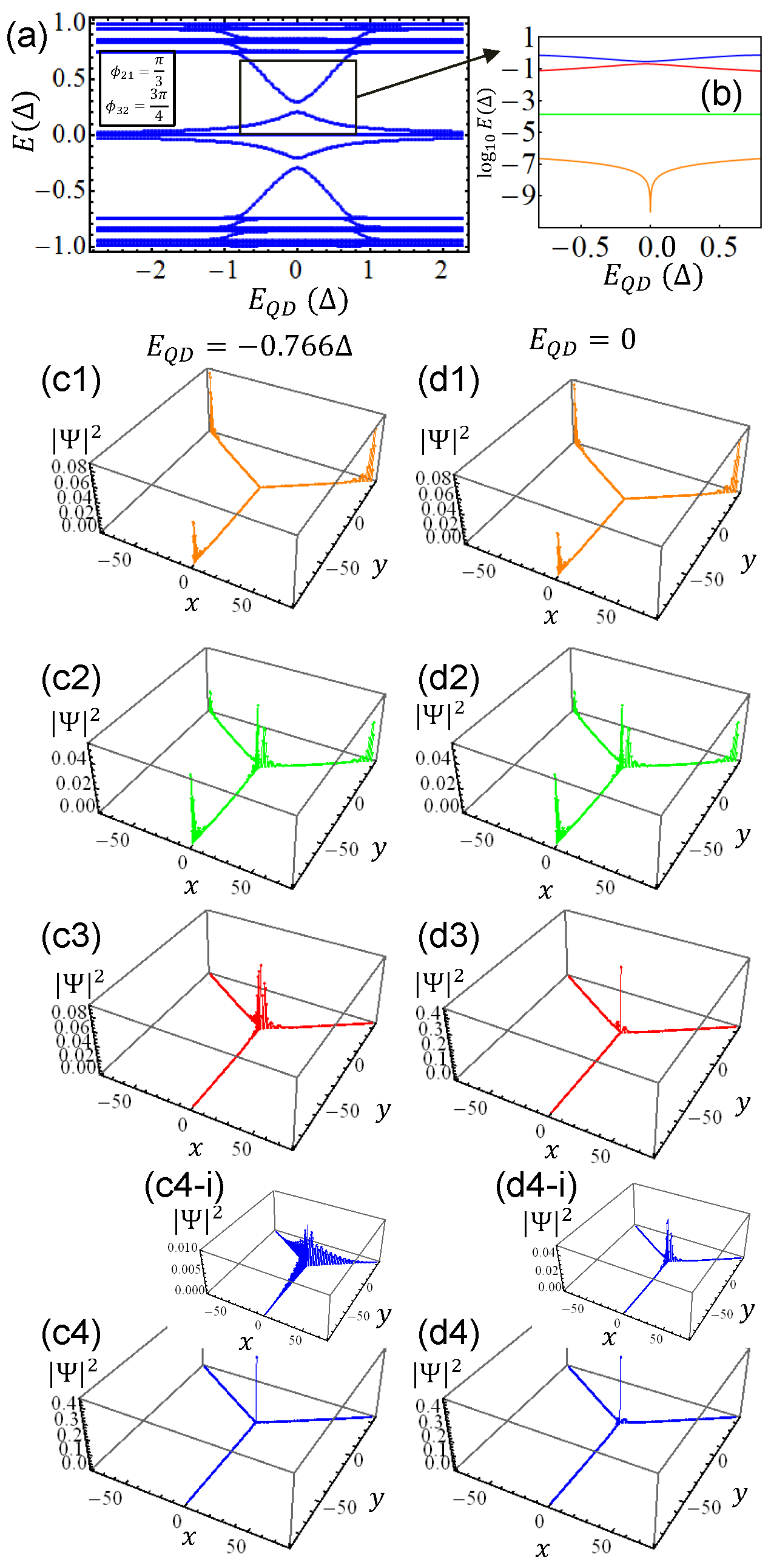}\protect\caption{\label{tightBindingEnWithResonantLevel}
Energy spectra and wave functions of subgap states of a three-terminal TSNW-QD hybrid system with a QD level near or inside the superconducting gap calculated based on the full BdG tight-binding Hamiltonian. (a) Energy spectra calculated for the system with phase differences $\phi_{21}=\frac{\pi}{3}$ and $\phi_{32}=\frac{3\pi}{4}$
against QD level energy $E_{Q\!D}$.  (b) Zoom-in view of the energy spectra of the four positive energy subgap states of the system in a logarithmic scale in the region marked by a rectangle in (a). (c1)-(c4) Probability distributions $|\Psi_{n}|^{2}$ of the four positive energy subgap states calculated for the system with $\phi_{21}=\frac{\pi}{3}$, $\phi_{32}=\frac{3\pi}{4}$ and $E_{Q\!D}=-0.766\Delta$. (c4-i) is the inset to (c4) which shows a zoom-in view of the probability distribution of the highest subgap state in the neighborhood of the QD.  (d1)-(d4) The same as (c1)-(c4) but for $E_{Q\!D}=0$. The couplings parameters employed in the calculations are $t_{ci}=\Delta$ (with $i=1,2,\mbox{and}\; 3$) and all the other unspecified parameters are the same as in Fig.~\ref{TBEnAndWaveFuncWithoutResonantLevel}.
}
\end{figure}

We now study the three-terminal TSNW-QD hybrid device in the case when a QD level is tuned into or near the superconducting gap. In the BdG particle-hole representation, a QD level contributes two states, namely the particle state and its hole partner, with opposite energies. Thus, there could totally exist eight subgap states in the system, instead of six seen in the previous subsection. However, we will show that this can only occur when the QD level is located near or inside the superconducting gap. Below, we will first discuss how the subgap states evolute with change in the energy position of the QD level, which can be tuned as
\begin{equation}
E_{Q\!D}=-eV_g+V_z+\delta_\Gamma,
\end{equation}
through a gate voltage $V_g$, where $\delta_\Gamma$ is the normalized energy of the QD level due to the couplings with the three TSNWs. Numerical calculation shows that $\delta_\Gamma\approx0.234\Delta$ in the system we have considered with coupling parameters $t_{ci}=\Delta$ in this subsection. Then, we will study the tuning of the subgap states by shifting the phase differences between the TSNWs.

Figure \ref{tightBindingEnWithResonantLevel} shows the subgap states of the three-terminal TSNW-QD hybrid system at given values of the phase differences between the TSNWs, $\phi_{21}$ and $\phi_{32}$, at different QD level energies. In Fig.~\ref{tightBindingEnWithResonantLevel}(a), the energy spectra calculated for the system with $\phi_{21}=\pi/3$ and $\phi_{32}=3\pi/4$ based on the full tight-binding Hamiltonian are plotted against the QD level energy ($E_{Q\!D}$). Figure \ref{tightBindingEnWithResonantLevel}(b) shows the energy positions in logarithmic scale of the positive energy subgap states in the region marked by a rectangle in Fig.~\ref{tightBindingEnWithResonantLevel}(a).
It is seen in Fig.~\ref{tightBindingEnWithResonantLevel}(a) that when the QD level is located far from the superconducting gap, only six subgap states are found in the gap. When the QD level is moved into the superconducting gap, a region, in which eight subgap states are present in the gap, exists [see Fig. \ref{tightBindingEnWithResonantLevel}(b) for the existence of the four positive energy subgap states]. When continuing moving the QD level away from the superconducting gap, the system hosts only six subgap states again and the results presented in the previous subsection are recovered.

Figures \ref{tightBindingEnWithResonantLevel}(c1) to \ref{tightBindingEnWithResonantLevel}(c4) show the wave functions of the four positive energy subgap states of the system at the QD level energy of $E_{Q\!D}=-0.766\Delta$, while Figs.~\ref{tightBindingEnWithResonantLevel}(d1) to \ref{tightBindingEnWithResonantLevel}(d4) show that at $E_{Q\!D}=0$.
It is seen in Fig.~\ref{tightBindingEnWithResonantLevel}(c1) and Fig.~\ref{tightBindingEnWithResonantLevel}(d1) that the lowest energy subgap state is built from a superposition of the three MBSs at the outer ends of the three TSNWs. Thus, the state is very close to zero in energy and does not show a significant change with shifting QD level. The second lowest subgap state is built from a superposition of all the six MBSs in the TSNWs and also does not show a significant change in energy with shifting QD level. The third and fourth lowest energy subgap states are built mainly from superpositions of the three MBSs adjacent to the QD and the QD state. Thus, the energies of these two states show clearly visible changes with shifting QD level position. However, the behaviors of these two subgap states are different. The highest subgap state moves towards a lower energy when the QD level moves towards the center of the superconducting gap, and is then moving back to higher energies and eventually to the outside of the superconducting gap with continuously increasing QD level energy. The wave function of this state is seen to be slightly more extended for the QD level at $E_{Q\!D}=-0.766\Delta$ [see Figs.~\ref{tightBindingEnWithResonantLevel}(c4-i)] than at $E_{Q\!D}=0$ [see Figs.~\ref{tightBindingEnWithResonantLevel}(d4-i)]. This is due to the fact that the subgap state is located deeper in energy in the superconducting gap at $E_{Q\!D}=0$ than at $E_{Q\!D}=-0.766\Delta$. In fact, this subgap state becomes more extended when the QD level is moved to a higher energy and eventually develops into an extended state when the QD level  merges into continuous band of quasiparticle states. On contrast, the third lowest subgap state moves to a higher energy when the QD level moves towards the center of the superconducting gap and then moves back to a lower energy after the QD level passes through the center of the gap. The wave function of this subgap state remains localized to the vicinity of the QD and does not show a significant change in localization with change in the QD level energy.

\begin{figure*}
\centering\includegraphics[scale=0.8]{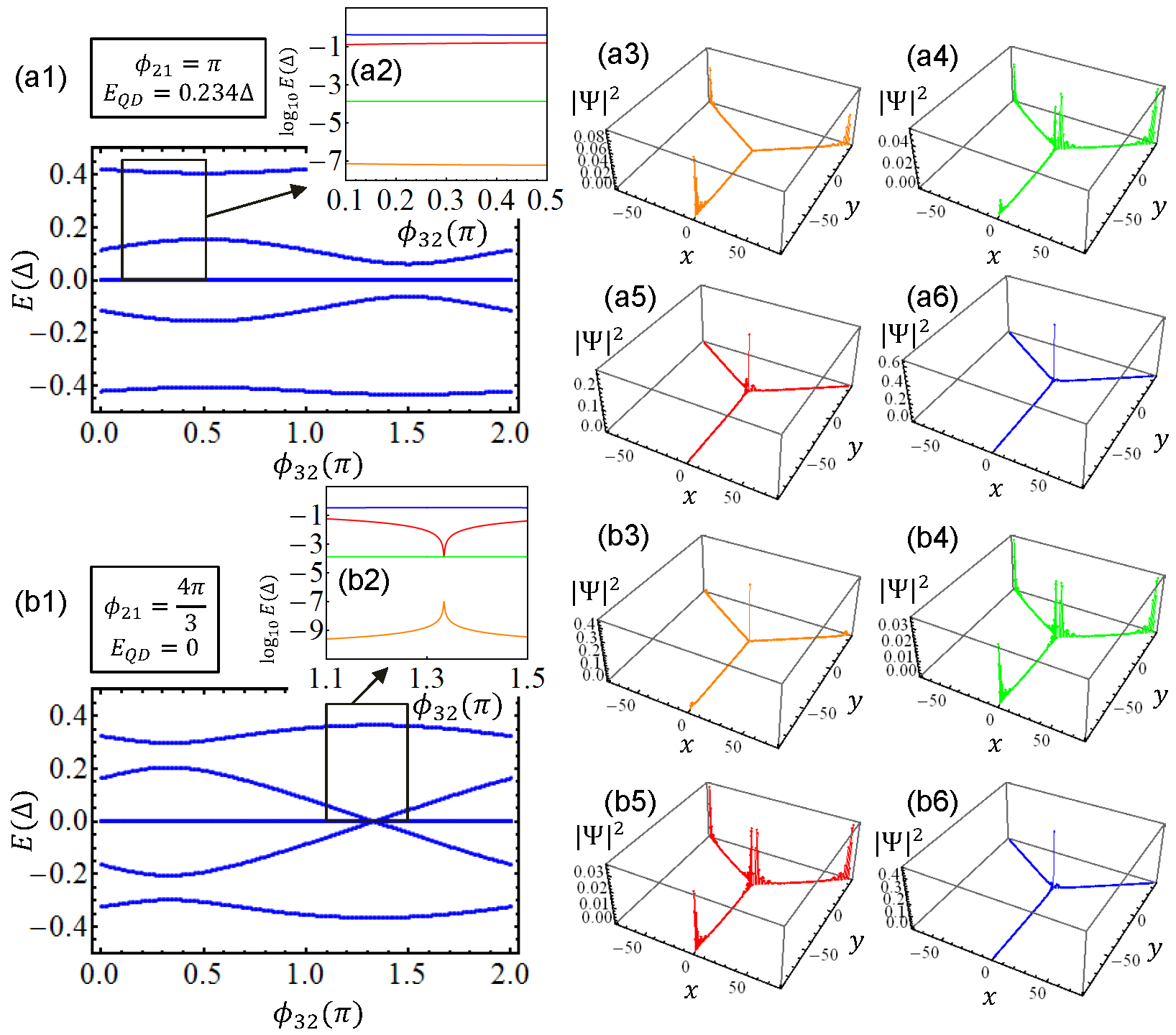}\protect\caption{\label{tightBindingEnWithFlux}
Subgap state energy spectra and wave functions of a three-terminal TSNW-QD hybrid system with a QD level near or inside the superconducting gap calculated based on the full BdG tight-binding Hamiltonian. (a1) Subgap state energy spectra calculated for the system with phase difference $\phi_{21}=\pi$ and QD level energy $E_{Q\!D}=0.234\Delta$ against phase difference $\phi_{32}$.  (a2) Zoom-in view of the energy spectra of the four positive energy subgap states of the system in a logarithmic scale in the region marked by a rectangle in (a1). (a3)-(a6) Probability distributions $|\Psi_{n}|^{2}$ of the four positive energy subgap states calculated for the system with $\phi_{21}=\pi$, $\phi_{32}=\frac{\pi}{3}$ and $E_{Q\!D}=0.234\Delta$. (b1)-(b2) The same as (a1)-(a2) but for $\phi_{21}=\frac{4\pi}{3}$ and $E_{QD}=0$. (b3)-(b6) The same as (a3)-(a6) but for $\phi_{21}=\phi_{32}=\frac{4\pi}{3}$ and $E_{QD}=0$. Other parameters are the same as in Fig.~\ref{tightBindingEnWithResonantLevel}}.
\end{figure*}

We now discuss the evolutions of the subgap states with changes in phase differences between TSNWs at different given QD level energies. Figure \ref{tightBindingEnWithFlux} shows the results of calculations based on the BdG tight-binding Hamiltonian for the system at the QD level energies of $E_{Q\!D}=0.234\Delta$ [Figs.~\ref{tightBindingEnWithFlux}(a1) to \ref{tightBindingEnWithFlux}(a6)] and $E_{Q\!D}=0$ [Figs.~\ref{tightBindingEnWithFlux}(b1) to \ref{tightBindingEnWithFlux}(b6)]. At both energies, the system hosts eight subgap states. it is seen in Figs.~\ref{tightBindingEnWithFlux}(a1) to \ref{tightBindingEnWithFlux}(a2), where $E_{QD}=0.234\Delta$ and $\phi_{21}=\pi$, that the energies of the eight subgap states only change slightly with varying $\phi_{32}$. The four positive energy subgap states are located at $\sim 10^{-7}\Delta$, $\sim 10^{-4}\Delta$, $\sim 10^{-1}\Delta$  and $\sim 10^{-1}\Delta$ in energy [see Fig.~\ref{tightBindingEnWithFlux}(a2)]. The wave functions of the four positive energy subgap states calculated at $\phi_{21}=\pi$ and $\phi_{32}=\pi/3$ are shown in Figs.~\ref{tightBindingEnWithFlux}(a3) to \ref{tightBindingEnWithFlux}(a6). It is seen in Fig.~\ref{tightBindingEnWithFlux}(a3) that the lowest positive energy subgap state at the energy of $\sim 10^{-7}\Delta$ is formed from the three MBSs located at the outer ends of the three TSNWs. The next lowest positive energy subgap state at the energy of $\sim 10^{-4}\Delta$ is formed from a superposition of all the six MBSs [see Fig.~\ref{tightBindingEnWithFlux}(a4)]. The reason for that the energies of these two subgap states are very close to zero but are still finite is the same as in the previous case, namely that the TSNWs are all finite in length. The probability distributions of the remaining two positive energy subgap states at the energies of $\sim 10^{-1}\Delta$ are shown in Figs.~\ref{tightBindingEnWithFlux}(a5) and \ref{tightBindingEnWithFlux}(a6). These two subgap states are formed from the hybridizations of the MBSs at the inner ends of the three TSNWs and the QD state. The probability distributions of all the eight subgap states vary little when $\phi_{21}$ and $\phi_{32}$ are changed, except for the case when $\phi_{21}$ and $\phi_{32}$ are in the vicinity of $\phi_{21}=4\pi/3$ and $\phi_{32}=4\pi/3$.

Figure \ref{tightBindingEnWithFlux}(b1) shows the evolution of the subgap state energy spectra at $E_{Q\!D}=0$ and $\phi_{21}=4\pi/3$ with change in $\phi_{32}$, while Fig.~\ref{tightBindingEnWithFlux}(b2) shows the energy evolution of the four lowest positive energy subgap states in logarithmic scale. It is seen that the two lowest positive energy subgap states at the energies of $\sim 10^{-9}\Delta$ and $\sim 10^{-4}\Delta$ are very close to zero and change little with change in phase difference $\phi_{32}$. The highest positive energy subgap state at the energy of $\sim 10^{-1}\Delta$ also exhibits a weak dependence on phase difference $\phi_{32}$.  The remaining, third lowest positive energy subgap state shows, however, a different behavior--it varies slowly in energy with increasing $\phi_{32}$ from zero and turns to move quickly towards smaller energy when $\phi_{32}$ approaches $4\pi/3$. The energy of this subgap state reaches the minimum of $\sim 10^{-4}\Delta$ at $\phi_{32}=4\pi/3$.
Figures \ref{tightBindingEnWithFlux}(b3) to \ref{tightBindingEnWithFlux}(b6) show the probability distributions of the four positive energy subgap states of the system at $\phi_{21}=\phi_{32}=4\pi/3$ and $E_{Q\!D}=0$. It is seen in Fig. \ref{tightBindingEnWithFlux}(b3) that the lowest positive energy subgap state at the energy of $\sim 10^{-9}\Delta$ is built dominantly from the QD state and the three MBSs at the outer ends of the three TSNWs. Note that three small but broad peaks are visible at the outer ends of the three TSNWs in Fig.~\ref{tightBindingEnWithFlux}(b3).
The second and third lowest positive energy subgap states at the energies of $\sim 10^{-4}\Delta$, seen in Figs.~\ref{tightBindingEnWithFlux}(b4) and \ref{tightBindingEnWithFlux}(b5), are formed dominantly from the MBSs in the three TSNWs. By a close examination, we also find that the probability distributions of these two subgap states in each TSNW have almost the same form of the probability distribution as in an isolated single TSNW, similar to the case discussed in the previous subsection. The  fourth lowest positive energy subgap state shown in Fig.~\ref{tightBindingEnWithFlux}(b6) is formed primarily from the QD state and the three MBSs adjacent to the QD, similar to the subgap state shown in the Fig.~\ref{tightBindingEnWithFlux}(a6).

\begin{figure*}
\centering\includegraphics[scale=0.8]{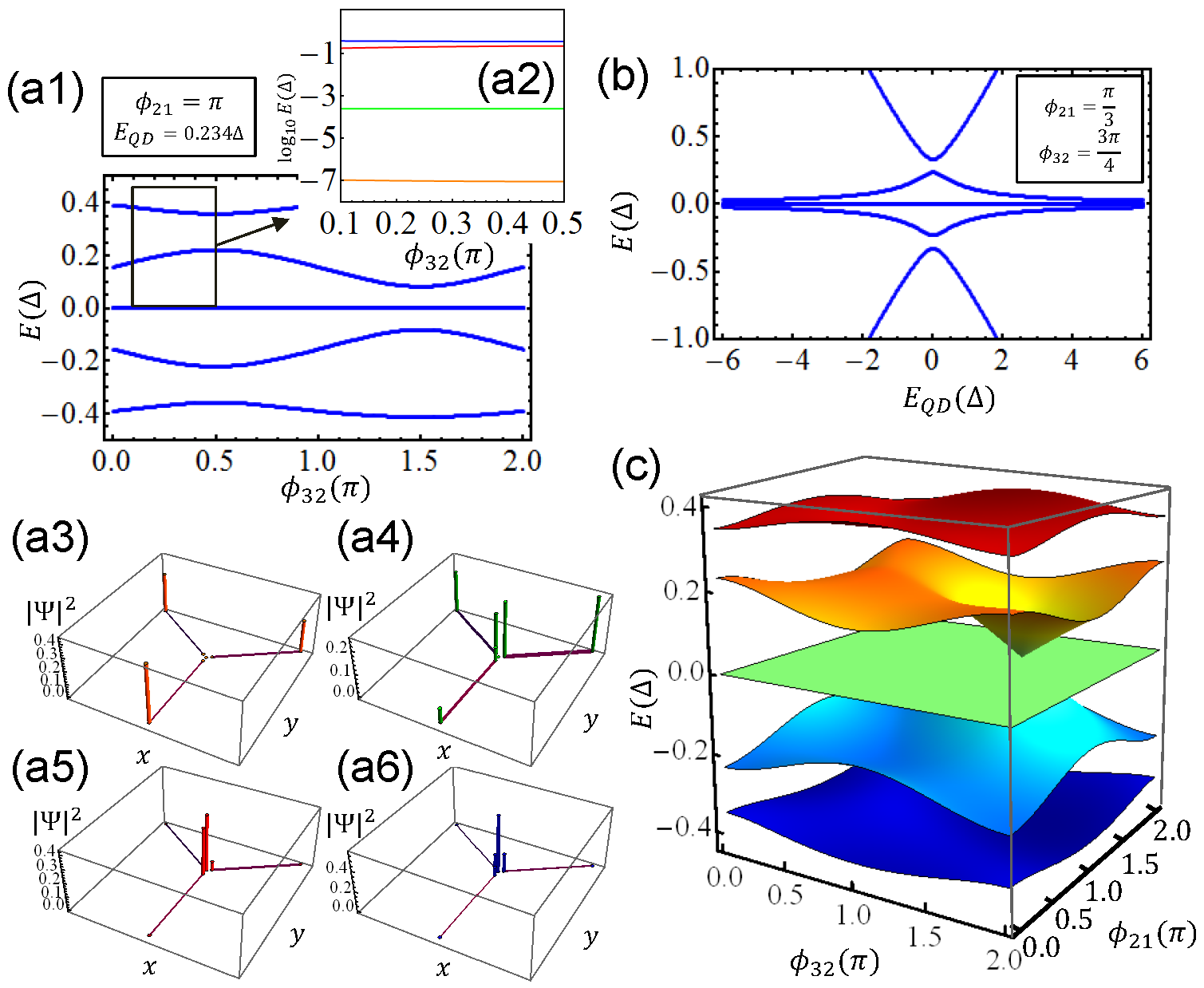}\protect\caption{\label{effectiveEnergyWithResonantLevel} Subgap state energy spectra and wave functions of a three-terminal TSNW-QD hybrid system with a QD level near or inside the superconducting gap calculated based on the effective model Hamiltonian of Eq.~(\ref{eqEffectiveQD}). (a1) Subgap state energy spectra calculated for the system with phase difference $\phi_{21}=\pi$ and QD level energy $E_{Q\!D}=0.234\Delta$ against phase difference $\phi_{32}$.  (a2) Zoom-in view of the energy spectra of the four positive energy subgap states of the system in a logarithmic scale in the region marked by a rectangle in (a1). (a3)-(a6) Probability distributions $|\Psi_{n}|^{2}$ of the four positive energy subgap states calculated for the system with $\phi_{21}=\pi$, $\phi_{32}=\frac{\pi}{3}$ and $E_{Q\!D}=0.234\Delta$. (b) Subgap state energy spectra calculated for the system with $\phi_{21}=\frac{\pi}{3}$ and $\phi_{32}=\frac{3\pi}{4}$ against QD level energy $E_{Q\!D}$. (c) Subgap state energy spectra of the system with $E_{Q\!D}=0.234\Delta$ in the full parameter space of $\phi_{32}$ and $\phi_{21}$. The coupling parameters employed in the calculations are $t_{14}=t_{25}=t_{36}=0.0005\Delta$ and $t_{Q\!Di}=0.33\Delta$ (with $i=1,2, \mbox{and}\; 3$).}
\end{figure*}

Based on the numerical results presented above, we repeat the procedure in the previous subsection to derive an effective model for describing the subgap states of the system by involving only the MBSs and the QD state as
\begin{equation}
H_{e\!f\!f}=H_{e\!f\!f}^{S}+H_{e\!f\!f}^{D}+H_{e\!f\!f}^{Q\!D},\label{eqEffectiveQD}
\end{equation}
where $H_{e\!f\!f}^{S}$ describes the couplings of the MBSs in the same TSNW,
$H_{e\!f\!f}^{D}$ is the couplings between the MBSs and the QD state, and $H_{e\!f\!f}^{Q\!D}$
is the Hamiltonian of the QD,
\begin{eqnarray}
H_{e\!f\!f}^{S} & = & i(t_{14}\gamma_{1}\gamma_{4}+t_{25}\gamma_{2}\gamma_{5}+t_{36}\gamma_{3}\gamma_{6}),\\
H_{e\!f\!f}^{D} & = & \sum_{i=1}^{3}t_{Q\!Di}(e^{i\frac{\beta_{i}}{2}}d^{\dagger}\gamma_{i}+\mathrm{H.c.}),\\
H_{e\!f\!f}^{Q\!D} & = & E_{Q\!D}d^{\dagger}d,
\end{eqnarray}
with the phases (not unique, up to a global phase background),
\begin{eqnarray}
\beta_{1} & = & q_{1},\nonumber \\
\beta_{2} & = & q_{2}+\phi_{21},\\
\beta_{3} & = & q_{3}+\phi_{21}+\phi_{32}.\nonumber
\end{eqnarray}
Similarly, the azimuths $q_{1,2,3}$ are in the range $[0,2\pi]$ for an arbitrary junction. Here, in the particular case shown in Fig.~\ref{setup}(b), $q_1=\frac{\pi}{6}$, $q_2=\frac{5\pi}{6}$, and $q_3=\frac{3\pi}{2}$, therefore
\begin{eqnarray}
\beta_{1} & = & \frac{\pi}{6},\nonumber \\
\beta_{2} & = & \frac{5\pi}{6}+\phi_{21},\\
\beta_{3} & = & \frac{3\pi}{2}+\phi_{21}+\phi_{32}.\nonumber
\end{eqnarray}
By solving the effective model Hamiltonian, the results of the BdG tight-binding calculations
can be reproduced. Figure \ref{effectiveEnergyWithResonantLevel} shows the results obtained based on the effective model Hamiltonian. When compared with Figs.~\ref{tightBindingEnWithFlux}(a1) to \ref{tightBindingEnWithFlux}(a6), it is seen that the energy spectra and the probability distributions are well recovered in the calculations based on the effective model Hamiltonian of Eq.~(\ref{eqEffectiveQD}).
Figure \ref{effectiveEnergyWithResonantLevel}(b) shows the energy spectra of the system at phase differences $\phi_{21}=\pi/3$ and $\phi_{32}=3\pi/4$ calculated against the QD level energy $E_{Q\!D}$ based the effective model Hamiltonian of Eq.~(\ref{eqEffectiveQD}). It is seen that the main features in the energy spectra of all the subgap states seen in Fig.~\ref{tightBindingEnWithResonantLevel}(a) are found in the calculations based on the effective model Hamiltonian. In Fig.~\ref{effectiveEnergyWithResonantLevel}(c), we show the energy spectra of the system with the QD level at $E_{Q\!D}=0.234\Delta$ in the whole parameter space of $\phi_{21}$ and $\phi_{32}$ based on the effective model Hamiltonian of Eq.~(\ref{eqEffectiveQD}).

For the TSNW-QD hybrid system with three semi-infinitely long TSNWs, $H_{e\!f\!f}^{S}$ can be neglected and the effective model Hamiltonian reads
\begin{eqnarray}
H_{S\!I\!n\!f} & = & H_{e\!f\!f}^{D}+H_{e\!f\!f}^{Q\!D}\label{eqEffectiveQDSemiInf}\nonumber\\
 & = & E_{Q\!D}d^{\dagger}d+\sum_{i=1}^{3}t_{Q\!Di}(e^{i\frac{\beta_{i}}{2}}d^{\dagger}\gamma_{i}+\mathrm{H.c.}).
\end{eqnarray}
Analytically solving the secular equation of the effective Hamiltonian in this infinitely long TSNW limit, we obtain eight subgap states with four zero-energy states and four finite-energy states. Similar to the case presented in the previous subsection, among the four zero-energy subgap states, three are the MBSs located at
the outer ends of the three TSNWs ($\gamma_{4}$, $\gamma_{5}$ and $\gamma_{6}$) and the remaining one is formed from a superposition
of the three inner MBSs adjacent to the QD as
\begin{equation}
E_{0}=0,\:\:
|E_{0}\rangle=\frac{1}{l_{N}}(l_{32}\gamma_1+l_{13}\gamma_2+l_{21}\gamma_3),
\end{equation}
where $l_{ij}\equiv t_{QDi}t_{QDj}\sin\frac{\beta_{ij}}{2}$, and
the normalized constant $l_{N}\equiv\sqrt{l_{21}^{2}+l_{32}^{2}+l_{13}^{2}}$.
The four finite-energy states are formed by hybridizations of the QD state and the three inner MBSs adjacent to the QD. Mathematically, these states are the solutions of a
quartic equation which can be simplified, by setting $x\equiv E^{2}$, to a quadratic equation,
\begin{equation}
x^{2}-Mx+N=0,
\end{equation}
where
\begin{eqnarray}
M=&\frac{E_{QD}^{2}+2(t_{QD1}^{2}+t_{QD2}^{2}+t_{QD3}^{2})}{4},
\nonumber\\
&N=\frac{l_{N}^{2}}{4}.\nonumber
\end{eqnarray}
Thus, the four finite-energy states have the eigenenergies of
\begin{equation}
E_{s_{1},s_{2}}=s_{1}\sqrt{\frac{M+s_{2}\sqrt{M^{2}-4N}}{2}},
\end{equation}
where $s_{1},s_{2}=\pm$. However, the eigenvectors $|E_{s_{1},s_{2}}\rangle$ of these four finite-energy states do not possess
a simple expression. Thus, it is more convenient to resort to numerical calculations for the eigenvectors of these states.

\subsection{Generality of the effective model Hamiltonians}

In the above subsections, the calculations based on the effective model Hamiltonians for the three-terminal TSNW-QD hybrid structures with the orientations of the three TSNWs set at $q_1=\frac{\pi}{6}$, $q_2=\frac{5\pi}{6}$, and $q_3=\frac{3\pi}{2}$ are presented and discussed.
However, the application of the effective model Hamiltonians to a three-terminal TSNW-QD hybrid structure with arbitrary TSNW orientations is straightforward. In fact,
the effective model Hamiltonians given by Eqs.~(\ref{eqEffectiveWL}) and (\ref{HSInf}) and by Eqs.~(\ref{eqEffectiveQD}) and (\ref{eqEffectiveQDSemiInf}) are in their general forms and can be readily used for the study of a three-terminal TSNW-QD hybrid structure with arbitrary TSNW orientations. For example, in the energy spectra shown in Figs.~\ref{energyEffective}(c) and \ref{effectiveEnergyWithResonantLevel}(c) obtained  based on the effective model Hamiltonians in the full parameter space of phase differences $\phi_{21}$ and $\phi_{32}$,  a specific point at which the six low energy subgap states are very close in energy (all in the orders of $\sim \pm10^{-4}\Delta$ or smaller). This point is seen to appear at $\phi_{21}=\phi_{32}=\frac{4\pi}{3}$ in Figs.~\ref{energyEffective}(c) and \ref{effectiveEnergyWithResonantLevel}(c). For a three-terminal TSNW-QD hybrid structure with arbitrary TSNW orientations, such a specific point always exists but shifts to
\begin{eqnarray}
\phi_{21} & = & 2\pi+q_{1}-q_{2},\\
\phi_{32} & = & 2\pi+q_{2}-q_{3}.
\end{eqnarray}
For a T-shaped three-terminal TSNW-QD hybrid structure \cite{np7.412}, for which $q_{1}=0$, $q_{2}=\pi$, and $q_{3}=\frac{3\pi}{2}$, the specific point is found to appear at $\phi_{21}=\pi$ and $\phi_{32}=\frac{3\pi}{2}$. For a parallel junction TSNW-QD hybrid system\cite{prb95.235305}, for which $q_{1}=0$, $q_{2}=\pi$, and
$q_{3}=2\pi$, the specific point shifts to $\phi_{21}=\phi_{32}=\pi$.
Thus, our effective model Hamiltonians given in Eqs.~(\ref{eqEffectiveWL}) and (\ref{HSInf}) and in Eqs.~(\ref{eqEffectiveQD}) and (\ref{eqEffectiveQDSemiInf}) are very general and can be applied to the calculations for a three-terminal TSNW-QD hybrid device with arbitrary nanowire orientations.

\section{Conclusions}

In conclusion, the subgap states in three-terminal TSNW-QD hybrid junction devices are studied.
The energy spectra and the wave functions  of the subgap states
are calculated as a function of the superconducting phase
differences between TSNWs and as a function of the QD level energy based on the BdG tight-binding Hamiltonians. In a system with no QD level located near or inside the superconducting band gap, there can exist six subgap states. Among them, two lowest energy (one positive and one negative energy) subgap states stay at nearly zero energies and are essentially formed by linear combinations of the three MBSs located at the outer ends of the three TSNWs. The next two lowest energy subgap states are built from linear combinations of all the six MBSs in the system and stay also at very small energies. However, the remaining two subgap states are mainly formed by linear combinations of the three MBSs adjacent to the QD and have significantly large finite energies. The calculations also show the existence of a specific point in the parameter space of phase differences between TSNWs, at which all the six subgap states move close to each other towards nearly zero energies. In a three-terminal TSNW-QD hybrid junction device with a QD level located near or inside the superconducting gap, there can appear eight subgap states. Among them, six low energy subgap states exhibit similar characteristics in energy positions and wave function localizations as in the system without a QD level located near or inside the superconducting gap except for the case where the QD level approaches zero energy. At this case, the two of the six subgap states at high energies contain significant contributions from the QD state. The remaining two large energy subgap states are the inherent property of the system with the QD level located near or inside the superconducting gap. These two states are constructed as linear combinations of the QD state and the three MBSs adjacent to the QD.

Simple but general effective model Hamiltonians for three-terminal TSNW-QD hybrid devices have also been extracted. The calculations based on the effective model Hamiltonians recover excellently the results of the calculations based on the full BdG tight-binding Hamiltonians. The effective model Hamiltonians can easily be generalized to the limit cases where the three TSNWs are  infinite in length. In such limit cases, some analytical solutions of the subgap states can be derived. The effective model Hamiltonians could be used as a starting point to construct and  investigate the braiding schemes of MBSs in TSNW junction devices.

\section*{Acknowledgments}

This work was supported by the Ministry of Science and Technology
of China (MOST) through the National Key Research and Development Program of
China (Grant Nos.~2016YFA0300601 and 2017YFA0303304),
the National Natural Science Foundation of China (Grant Nos.~91221202, 91421303, and 11604005), and the Swedish
Research Council (VR). GYH would also like to acknowledge financial
support from the China Postdoctoral Science Foundation (Grant No.~2016M591001)
to this work.

\end{document}